\documentclass{elsart}

\usepackage{graphicx}
\usepackage{ifthen}
\usepackage{xspace}

\usepackage{cite}                 

\newcommand{\sczmdiff}{\ensuremath{167.38}\xspace}
\newcommand{\sczmstat}{\ensuremath{0.21}\xspace}
\newcommand{\sczmsyst}{\ensuremath{0.13}\xspace}

\newcommand{\scdmdiff}{\ensuremath{167.35}\xspace}
\newcommand{\scdmstat}{\ensuremath{0.19}\xspace}
\newcommand{\scdmsyst}{\ensuremath{0.12}\xspace}

\newcommand{\scmdiff}{\ensuremath{-0.03}\xspace}
\newcommand{\scmstat}{\ensuremath{0.28}\xspace}
\newcommand{\scmsyst}{\ensuremath{0.11}\xspace}

\newcommand{\sczyield}{\ensuremath{362\pm36}\xspace}
\newcommand{\scdyield}{\ensuremath{461\pm39}\xspace}

\newcommand{\cer}{\v{C}erenkov\xspace}

\newcommand{\mc}{Monte Carlo\xspace}

\newcommand{\mathsl}[1]{\mbox{\textsl{#1}}}

\newcommand{\meson}[1]{\ensuremath{\mathsl{#1}}}
\newcommand{\baryon}[1]{\ensuremath{\mathsl{#1}}}

\newcommand{\quark}[1]{\ensuremath{\mathsl{#1}}}

\newcommand{\uq}{\quark{u}\xspace}

\newcommand{\dq}{\quark{d}\xspace}

\newcommand{\proton}{\baryon{p}\xspace}

\newcommand{\pion}{\ensuremath{\pi}\xspace}
\newcommand{\piplus}{\ensuremath{\pion^+}\xspace}
\newcommand{\piminus}{\ensuremath{\pion^-}\xspace}

\newcommand{\kaon}{\ensuremath{\meson{K}}\xspace}

\newcommand{\dmeson}{\ensuremath{\meson{D}}\xspace}
\newcommand{\dzero}{\ensuremath{\dmeson^{0}}\xspace}
\newcommand{\dsp}{\ensuremath{\dmeson^{*+}}\xspace}

\newcommand{\lc}{\ensuremath{\Lambda_c^+}\xspace}

\newcommand{\sigc}{\ensuremath{\Sigma_c}\xspace}
\newcommand{\scd}{\ensuremath{\Sigma_c^{++}}\xspace}

\newcommand{\scz}{\ensuremath{\Sigma_c^{0}}\xspace}

\newcommand{\lcstaru}{\ensuremath{\Lambda_{c1}^{*+}(2625)}\xspace}

\newcommand{\pkpi}{\ensuremath{\lc \to \proton \kaon^- \piplus}\xspace}

\newcommand{\lcpip}{\ensuremath{\scd \to \lc \piplus}\xspace}

\newcommand{\lcpim}{\ensuremath{\scz \to \lc \piminus}\xspace}

\newcommand{\mdscdscz}{\ensuremath{M(\scd -  \scz)}\xspace}
\newcommand{\mdscdlc}{\ensuremath{M(\scd -  \lc)}\xspace}
\newcommand{\mdsczlc}{\ensuremath{M(\scz -  \lc)}\xspace}

\newcommand{\mevcc}{\ensuremath{\mathrm{MeV}/c^2}\xspace}

\newcommand{\gevc}{\ensuremath{\mathrm{GeV}/c}\xspace}
\newcommand{\gev}{\ensuremath{\mathrm{GeV}}\xspace}

\newcommand{\eqnref}[1]{Equation~(\ref{eqn:#1})}
\newcommand{\figref}[1]{Figure~\ref{fig:#1}}
\newcommand{\tabref}[1]{Table~\ref{tab:#1}}

\newcommand{\figlabel}[1]{\label{fig:#1}}
\newcommand{\eqnlabel}[1]{\label{eqn:#1}}
\newcommand{\tablabel}[1]{\label{tab:#1}}

\graphicspath{{sigmac_mass/}}

\begin{document}                                                 

\begin{frontmatter}
\title{
Measurements of  the \\ \scz and \scd Mass Splittings
}

The FOCUS Collaboration

\author[davis]{J.M.~Link},
\author[davis]{V.S.~Paolone\thanksref{atpitt}},
\author[davis]{M.~Reyes\thanksref{atmex}},
\author[davis]{P.M.~Yager},
\author[cbpf]{J.C.~Anjos},
\author[cbpf]{I.~Bediaga},
\author[cbpf]{C.~G\"obel\thanksref{aturug}},
\author[cbpf]{J.~Magnin\thanksref{atbogota}},
\author[cbpf]{J.M.~de~Miranda},
\author[cbpf]{I.M.~Pepe\thanksref{atbahia}},
% ...

\address[davis]{University of California, Davis, CA 95616}
\address[cbpf]{Centro Brasileiro de Pesquisas F\'\i sicas, Rio de Janeiro,
RJ, Brazil}
\address[cine]{CINVESTAV, 07000 M\'exico City, DF, Mexico}
\address[cu]{University of Colorado, Boulder, CO 80309}
\address[fnal]{Fermi National Accelerator Laboratory, Batavia, IL 60510}
\address[fras]{Laboratori  Nazionali di Frascati dell'INFN, Frascati, Italy,
      I-00044}
\address[ui]{University of Illinois, Urbana-Champaign, IL 61801}
\address[iu]{Indiana University, Bloomington, IN 47405}
\address[koru]{Korea University, Seoul, Korea 136-701}
\address[milan]{INFN and University of Milano, Milano, Italy}
\address[anc]{University of North Carolina, Asheville, NC 28804}
\address[pavia]{Dipartimento di Fisica Nucleare e Teorica and INFN,
Pavia, Italy}
\address[pr]{University of Puerto Rico, Mayaguez, PR 00681}
\address[sc]{University of South Carolina, Columbia, SC 29208}
\address[ut]{University of Tennessee, Knoxville, TN 37996}
\address[vandy]{Vanderbilt University, Nashville, TN 37235}
\address[wisc]{University of Wisconsin, Madison, WI 53706}
\address[yon]{Yonsei University, Seoul, Korea 120-749}

\thanks[atpitt]{Present Address: University of Pittsburgh, Pittsburgh,
PA 15260}
\thanks[atmex]{Present Address: Instituto de F\'\i sica y
Matematicas, Universidad Michoacana de
San Nicolas de Hidalgo, Morelia, Mich., M\'exico 58040}
\thanks[aturug]{Present Address: Instituto de F\'\i sica, Facultad de
Ingenier\'\i a, Univ. de la Rep\'ublica, Montevideo, Uruguay}
\thanks[atbogota]{Present Address: Universidad de los Andes, Bogota,
Colombia}
\thanks[atbahia]{Present Address: Instituto de F\'\i sica, Universidade
Federal da Bahia, Salvador, Brazil}
\thanks[atmilan]{Present Address: INFN sezione di Milano, Milano,
Italy}
\thanks[atpr]{Present Address: University of Puerto Rico, Mayaguez,
PR  00681}
\thanks[atpub]{Present Address: Instituto de F\'{\i}sica, Universidad
Aut\'onoma de Puebla,  Puebla, M\'exico}
\thanks[atncar]{Present Address: National Center for Atmospheric Research, Boulder, CO} 
\thanks[atlbl]{Present Address: Lawrence Berkeley Lab, Berkeley, CA
94720}
\thanks[atbrescia]{Present Address:
Dipartimento di Chimica e F\'\i sica per l'Ingegneria e per i Materiali,
Universita' di Brescia  and  INFN sezione di Pavia}
\thanks[atromania]{Present Address: Nat. Inst. of Phys. and Nucl. Eng., Bucharest, Romania}
\thanks[atcu]{Present Address: University of Colorado, Boulder, CO 80309}
\thanks[atlucent]{Present Address: Lucent Technology}
\thanks[atagusta]{Present Address: Augusta Technical Inst., Augusta, GA
30906}
\thanks[atadams]{Present Address: Adams State College, Alamosa, CO 81102}

\begin{abstract}   
Using a high statistics sample of photoproduced charmed particles from the
\mbox{FOCUS} experiment at Fermilab (FNAL-E831), we measure the mass splittings
of the charmed baryons \scz and \scd.  We find $\mdsczlc = \sczmdiff \pm
\sczmstat \pm \sczmsyst~\mevcc$ and $\mdscdlc = \scdmdiff \pm \scdmstat \pm
\scdmsyst~\mevcc$ with samples of \sczyield and \scdyield events,
respectively.  We measure the isospin mass splitting $\mdscdscz$ to be
$\scmdiff \pm \scmstat \pm \scmsyst~\mevcc$. The first errors are statistical
and the second are systematic.
\end{abstract}
\end{frontmatter}

Many
experiments~\cite{Aitala:1996cy,Crawford:1993pv,pdg:frabetti96,Aleev:1996ym,Albrecht:1988gc,Anjos:1989mk,Diesburg:1987kx,Bowcock:1989qh}
have measured the mass differences of the \scz and \scd baryons with respect to
the \lc.  Only FNAL-E791~\cite{Aitala:1996cy} and
CLEO~II~\cite{Crawford:1993pv} have measured the mass differences with respect
to the \lc to a total (statistical and systematic) precision of less than
$0.5~\mevcc$.  Some of these previous measurements have suggested that the
\sigc multiplet is unique in that the masses of the isospin states
\emph{increase} with the quark substitution $\dq \to \uq$. Such a result is not
at odds with theoretical calculations since there are several canceling
terms necessary to calculate the hyperfine mass splittings.  In addition to the
constituent quark mass differences, effects from the potential model used as
well as the Coulomb interaction and hyperfine interactions must also be
considered~\cite{theory:rosner_ms_98}.

In this paper, we present a measurement using data from the FOCUS experiment
which improves upon the earlier measurements and confirms that the \sigc
isospin mass splitting is much smaller than for other baryon isospin
multiplets. FOCUS collected data using the Wideband photon beamline during the
1996--1997 Fermilab fixed-target run and is an upgraded version of
FNAL-E687~\cite{Frabetti:1992au}.  The FOCUS experiment utilizes a forward
multiparticle spectrometer to study the interactions of high energy photons 
($\langle E \rangle \approx 180~\gev$) with a segmented BeO target.  

Charged particles are tracked within the spectrometer by two silicon
microvertex detector systems. One system is interleaved with the target
segments; the other is downstream of the target region. These detectors provide
excellent separation of the production and decay vertices. Further downstream,
charged particles are tracked and momentum analyzed by a system of five
multiwire proportional chambers and two dipole magnets with opposite polarity.
Three multicell threshold \cer detectors are used to discriminate between
electrons, pions, kaons, and protons.

To reconstruct the decays $\sigc \to \lc \pi^\pm$, we first obtain a sample of
\lc baryons\footnote{Throughout this paper, charge conjugate states are
implicitly included unless stated otherwise.} using the decay mode \pkpi.
During its run, FOCUS collected in excess of 25\,000 fully reconstructed decays
in the channel \pkpi from $6\times 10^9$ hadronic triggers. Potential \lc
candidates are distinguished from background hadronic interactions primarily by
requiring that the production and decay vertices are distinct. We apply a
minimum detachment requirement of 6, which requires that the measured
separation of the two vertices divided by the error on that measurement is
greater than 6. We also ensure that both vertices are well formed by requiring
a confidence level greater than 1\% on the fit to each vertex.

The \pkpi decay channel is separated from other three body decays that
reconstruct with masses in the \lc mass window by applying \cer based particle
identification to the daughter particles. The \cer particle identification cuts
used in \mbox{FOCUS} are based on likelihood ratios between the various stable
particle identification hypotheses. The product of all firing probabilities for
all cells within the \cer cones in each detector produces a $\chi^2$-like
variable  $W_i = -2 \ln (\mathrm{likelihood})$ where $i$ ranges over the
electron, pion, kaon, and proton hypotheses.

Tight cuts are placed on the proton candidate particle requiring that the
proton hypothesis is favored over both the pion and kaon hypotheses. We require 
that $W_{\pion} - W_{p} > 4$ and $W_{K} - W_{p} > 1$.
For the kaon candidate, the kaon hypothesis is required to be favored over the
pion hypothesis by requiring $W_{\pion} - W_{K} > 3$. For the pion
candidate, we require that no hypothesis is favored over the pion hypothesis
with a $\Delta W$ exceeding 6.

The \lc candidates are required to have a momentum greater than 40~\gevc and a
proper decay time less than 10 times the mean lifetime of the \lc.  This final
requirement is effective in removing background contamination from the longer
lived charm mesons. The invariant mass distribution for \lc candidates which
satisfy all the selection criteria is shown in \figref{sigmac_lc_mass}.

\begin{figure}
\begin{center}
{\includegraphics[width=9cm]{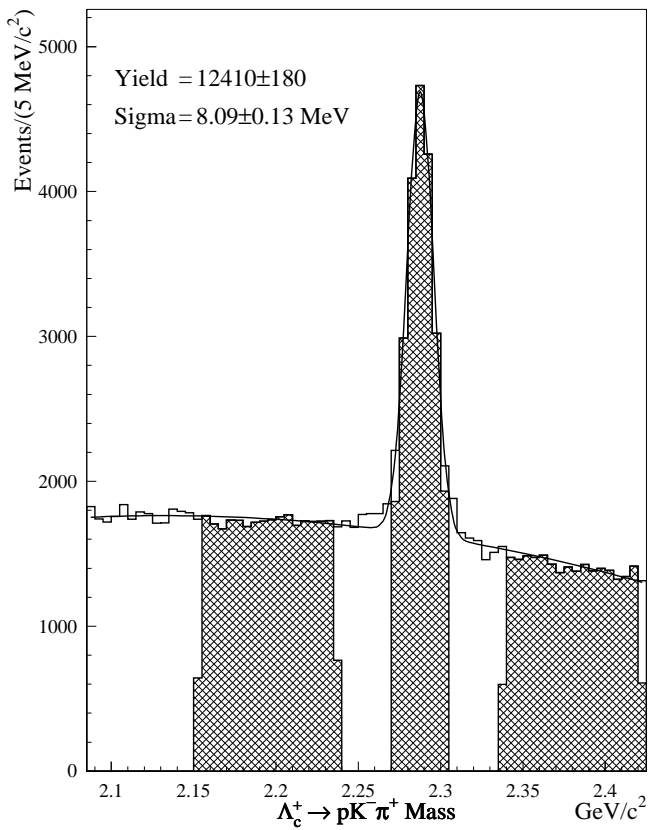}}
\end{center}
\caption[\lc candidates used in the reconstruction of \scz and \scd]{\pkpi
candidates used in the reconstruction of \scz and \scd candidates.  The central
hatched region illustrates the cut around the nominal \lc mass.  The
outer hatched sidebands are used in background studies. The mass cut and bin
boundaries do not coincide.}
\figlabel{sigmac_lc_mass}
\end{figure}

\sigc candidates are reconstructed by combining the \lc candidates within $2.1
\sigma$ of the mean \lc mass with a charged pion.\footnote{Referred to as a
``soft pion'' since it is usually low momentum.}  As before, a vertex with
confidence level greater than 1\% is required between the \lc candidate and the
pion. For the soft pion candidate, we require that no hypothesis is favored
over the pion hypothesis with a $\Delta W$ exceeding 4. We obtain a sample of
\sczyield \lcpim and \scdyield \lcpip decays.

To remove any systematic effects due to the reconstruction of the \lc,
we compute and plot the invariant mass difference. The computed \lc momentum
and mass are combined with the pion momentum and known mass to form
$M(\sigc)$.  The computed \lc mass is subtracted to obtain $\Delta M$.     

The resulting distributions are then fit with the background function
\begin{equation}
N(1 + \alpha(\Delta M-m_\pi)\Delta M^{\beta})
\eqnlabel{free_bg}
\end{equation}
where $N$, $\alpha$, and $\beta$ are allowed to vary. A Gaussian fitting
function is used to represent the \sigc signals. A small component (described
below) attributed to $\lcstaru \to \lc \piplus \piminus$ decays is also
included. The invariant mass distributions, fits, and fit values are shown in
\figref{sigmac_mass}. Backgrounds in both invariant mass plots vary smoothly
across the region of the mass peak and in a similar fashion to the data outside
the peak. Thus, the use of a smoothly varying function such as \eqnref{free_bg}
is justified.

\begin{figure}[tbp]
\begin{center}
{\includegraphics[width=12cm,bb=20 250 580 740]{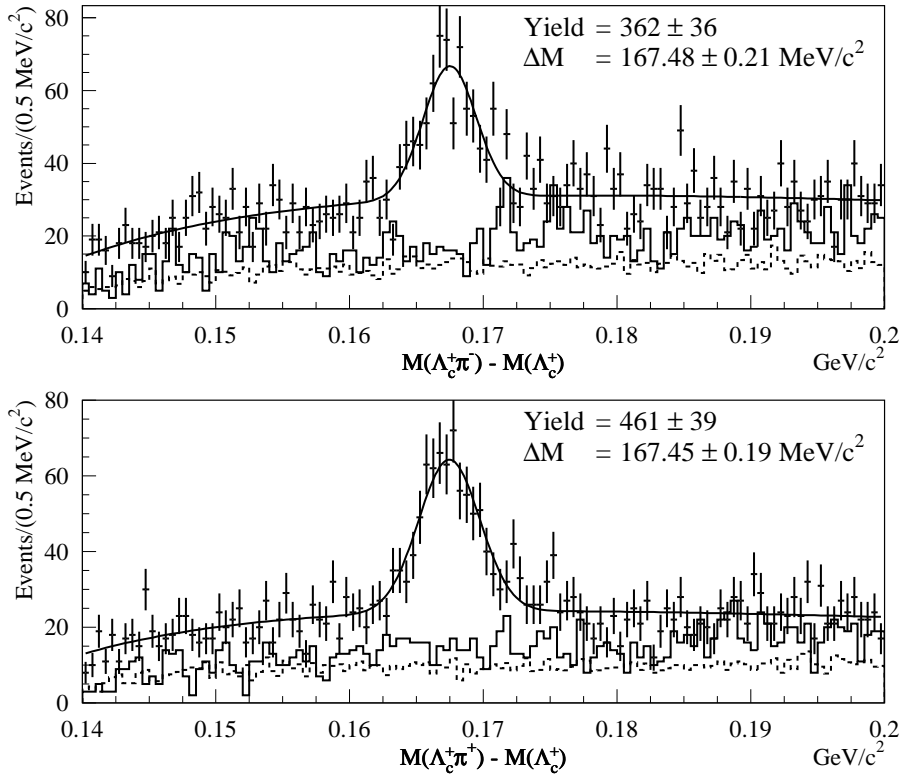}}
\end{center}
\caption[Mass difference distributions for \mdsczlc and \mdscdlc]{Mass
difference distributions for \mdsczlc and \mdscdlc. The dotted histograms are
from the sidebands in \figref{sigmac_lc_mass}.  The lower solid histograms are
those formed by combining \lc candidates with pions from the previous event
containing a \lc candidate. These mass difference values have not been
corrected with the mass calibration adjustment described in the text.}
\figlabel{sigmac_mass}
\end{figure}

Several sources of systematic error were investigated, including
knowledge of our momentum scale, reconstruction and fitting biases, and biases
in the analysis cuts.  Systematic errors on the three measured quantities are
calculated separately since systematic effects on the value of \mdscdscz are
expected to be smaller than those on the measurements of the $\sigc - \lc$ mass
differences.

From studies of the reconstructed \dmeson masses and the decay $\dsp \to \dzero
\piplus$ we estimate that the measured $\sigc-\lc$ mass differences are
$0.10\pm0.05$~\mevcc above the true values.  The final values quoted are
adjusted for this shift and a systematic uncertainty of $0.05$~\mevcc is
incorporated into the systematic error. Our measurement of the $\scd-\scz$
mass difference is unchanged by this shift. We attribute this effect to a
slight momentum miscalibration.

We find a maximum systematic shift of $\pm 0.04~\mevcc$ on the measurements of \mdsczlc
and \mdscdlc when we reconstruct \mc events and compare the measured values  to
the input values.  No uncertainty on \mdscdscz is found from this source. These
shifts are included as systematic errors.

We also vary the fitting and reconstruction methods and assess the effect on
the final measurement.  We change the background description to two components
of the functional form of \eqnref{free_bg}.  The shapes of the two components
are derived from using \lc candidates from the mass sidebands (shown in
\figref{sigmac_lc_mass}) and from combinations of \lc candidates from one event
with pions from a different event.  Both distributions are shown in
\figref{sigmac_mass}. The shapes of these distributions are fixed and the
normalizations are allowed to vary.   The effect of using an alternate
background function given by
\begin{equation}
A + B \sqrt{\Delta M^2 - m_{\pion}^2} + C\cdot\Delta M
\eqnlabel{cleo_func}
\end{equation}
where $A$, $B$, and $C$ are allowed to vary, has also been studied.

By default, we include the contribution from the decay $\lcstaru \to \lc
\piplus \piminus$. The magnitude of this contribution is obtained from a \mc
simulation normalized to the number of reconstructed $\lcstaru \to \lc \piplus
\piminus$ decays. This contribution accounts for approximately 500 events in
each of the $\lc \pi^{\pm}$ backgrounds. The effect of excluding this feed-down
contribution was also studied and found to be minimal.   

Finally, we measure the \sigc mass differences using direction vectors obtained
from the ``downstream'' silicon detector rather than those obtained by
combining information from both silicon detectors. The systematic errors
obtained in these variations range from 0.02--0.06~\mevcc, depending on the
measurement.

The final systematic checks are performed using a ``split sample'' technique to
estimate systematic errors. In this technique, we divide the data into two
roughly equal portions based on kinematic variables or running conditions and
perform the measurement on each statistically independent subsample. We choose
variables where we might expect, either through reconstruction methods or
changes in running conditions, to introduce a bias in the measured quantity. By
using a method similar to the $S$-factor method of the Particle Data Group
\cite[pg. 10]{pdg:pdg98}, an attempt is made to consider only systematic
effects which arise from true differences in the measured values, rather than
from the expected statistical fluctuations. We split the data based on
particle/anti-particle,\footnote{This check is especially important since any
difference in the reconstruction of positively and negatively charged soft
pions could introduce a systematic error into the measurement of \mdscdscz.}
detachment cut, \lc momentum, soft pion momentum, and into two running periods,
one before and one after the installation of the interleaved silicon system. 
We estimate the combined split sample systematic error to be 0.10~\mevcc for
the $M(\sigc - \lc)$ measurements as well as for the $M(\scd-\scz)$
measurement.

As an additional check on our ability to reconstruct excited charm states in an
unbiased manner, many of the same systematic effects were studied for the decay
$\dsp \to \dzero \piplus$ using a significant portion of the FOCUS data sample.
This mode is not statistically limited and allows for the detection of very
small systematic effects. No such effects of consequence were discovered.

The systematic errors on these measurements are summarized and totaled in
\tabref{syst_summary}. The totals are determined by adding the various errors in
quadrature.    

\begin{table}
\centering
\caption[Systematic errors for  \scd and \scz mass differences]{Systematic errors for  \scd and \scz mass differences.}
\tablabel{syst_summary}
\begin{tabular}{lccc} 
\hline 
                                 & \multicolumn{3}{c}{Error (\mevcc)}  \\ \cline{2-4}
\multicolumn{1}{c}{Source} & $\scd - \lc$ & $\scz - \lc$ & $\scd - \scz$  \\ \hline
$p$ scale                    & $0.05$       & $0.05$       & $0.00$ \\
Bias from MC                 & $0.04$       & $0.04$       & $0.00$ \\
Fitting                      & $0.02$       & $0.06$       & $0.04$ \\
Split samples                & $0.10$       & $0.10$       & $0.10$ \\
\hline
Total                        & $0.12$       & $0.13$       & $0.11$ \\
\hline
\end{tabular} 
\end{table}

Considering both the statistical and systematic errors and applying the shift
due to the momentum miscalibration, we find final values of $\mdsczlc =
\sczmdiff \pm \sczmstat \pm \sczmsyst~\mevcc$, $\mdscdlc = \scdmdiff \pm
\scdmstat \pm \scdmsyst~\mevcc$, and $\mdscdscz = \scmdiff \pm \scmstat \pm
\scmsyst~\mevcc$, where the first errors are statistical and the second are
systematic.  Our values are compared with the two previous
measurements of comparable precision in \tabref{sc_mass_comp}.

\begin{table}
\centering
\caption[Comparison of measurements of \scd and \scz mass differences]{Comparison of measurements of \scd and \scz mass differences.}
\tablabel{sc_mass_comp}
\begin{tabular}{llll} 
\hline 
                                 & \multicolumn{3}{c}{Mass difference (\mevcc)}  \\ \cline{2-4}
\multicolumn{1}{c}{Experiment} & \multicolumn{1}{c}{$\scd - \lc$} & \multicolumn{1}{c}{$\scz - \lc$}&
\multicolumn{1}{c}{$\scd - \scz$}  \\ \hline
CLEO~II~\cite{Crawford:1993pv} & $168.20 \pm 0.30 \pm 0.20$ & $167.10 \pm 0.30 \pm 0.20$ & $+1.10 \pm 0.40 \pm 0.10$ \\
E791~\cite{Aitala:1996cy}      & $167.76 \pm 0.29 \pm 0.15$ & $167.38 \pm 0.29 \pm 0.15$ & $+0.38 \pm 0.40 \pm 0.15$  \\
World avg.~\cite{pdg:pdg98}    & $167.87 \pm 0.20$          & $167.31 \pm 0.21$          & $+0.66 \pm 0.28$ \\
This work                      & $\scdmdiff \pm \scdmstat \pm \scdmsyst$ & $\sczmdiff \pm \sczmstat \pm \sczmsyst$ &$\scmdiff \pm \scmstat \pm \scmsyst$   \\
\hline
\end{tabular} 
\end{table}

Our measurement does not support the conclusion that the \scd is more massive
than the \scz; rather our measurement of the mass difference between the \scd
and \scz is consistent with zero. However, for all other well measured baryons
the corresponding difference is positive. The \sigc multiplet does appear to be
unique in this regard.

A number of theoretical calculations of \mdscdscz are presented in
\tabref{sum_predict}; many are excluded by the present measurement.
(The calculations presented in Varga, \etal\ are recent recalculations of a
number of earlier predictions.) 

\begin{table}
\centering
\caption[Theoretical predictions of \mdscdscz]{Theoretical predictions of \mdscdscz.}
\tablabel{sum_predict}
\begin{tabular}{lcc} 
\hline 
\multicolumn{1}{c}{Author}    & \mdscdscz (\mevcc)  \\ \hline
Capstick~\cite{Capstick:1987cw} & 1.4      \\
Chan~\cite{Chan:1985ty}         & 0.3      \\
Hwang~\cite{Hwang:1987ee}       & 3.0      \\
Isgur~\cite{Isgur:1980ed}       & $-2.0$      \\
Richard~\cite{Richard:1984}    & $-2$ or $3$   \\
Sinha~\cite{Sinha:1989}      & $1.5 \pm 0.2$   \\
Varga~\cite{Varga:1999wp}    & $-6$ to 18   \\
\hline
\end{tabular} 
\end{table}

In conclusion, we report new measurements of the quantities \mdsczlc,
\mdscdlc, and \mdscdscz which represent significant improvements on the
world's best measurements. We find no evidence for an isospin mass splitting 
between the \scz and \scd baryons.

We wish to acknowledge the assistance of the staffs of Fermi National
Accelerator Laboratory, the INFN of Italy, and the physics departments of the
collaborating institutions. This research was supported in part by the U.~S.
National Science Foundation, the U.~S. Department of Energy, the Italian
Istituto Nazionale di Fisica Nucleare and Ministero dell'Universit\`a e della
Ricerca Scientifica e Tecnologica, the Brazilian Conselho Nacional de
Desenvolvimento Cient\'{\i}fico e Tecnol\'ogico, CONACyT-M\'exico, the Korean
Ministry of Education, and the Korean Science and Engineering Foundation.

\bibliographystyle{phaip}
\bibliography{sigmac_mass_paper}

\end{document}